\documentclass{article} 
\usepackage{iclr_conference,times,natbib}

\usepackage{amsmath,amsfonts,bm}

\def\eqref#1{equation~\ref{#1}}

\def\1{\bm{1}}

\DeclareMathAlphabet{\mathsfit}{\encodingdefault}{\sfdefault}{m}{sl}
\SetMathAlphabet{\mathsfit}{bold}{\encodingdefault}{\sfdefault}{bx}{n}

\usepackage{enumitem,amssymb}

\usepackage{hyperref}
\usepackage{url}
\usepackage{graphicx}

\title{RESuM: Rare Event Surrogate Model for  Physics Detector Design}

\author{Ann-Kathrin Schuetz \\
Nuclear Science Division \\
Lawrence Berkeley National Laboratory \\
Berkeley, CA 94720 \\
\texttt{aschuetz@lbl.gov} \\
\And
Alan Poon \\
Nuclear Science Division \\
Lawrence Berkeley National Laboratory \\
Berkeley, CA 94720 \\
\texttt{awpoon@lbl.gov} \\
\And
Aobo Li \\
Halıcıoğlu Data Science Institute\\
Department of Physics\\
University of California San Diego \\
La Jolla, CA  92093\\
\texttt{liaobo77@ucsd.edu} \\
}

\iclrfinalcopy 
\begin{document}

\maketitle

\begin{abstract}

The experimental discovery of neutrinoless double-beta decay (NLDBD) would answer one of the most important questions in physics: Why is there more matter than antimatter in our universe? To maximize the chances of detection, NLDBD experiments must optimize their detector designs to minimize the probability of background events contaminating the detector. Given that this probability is inherently low, design optimization either requires extremely costly simulations to generate sufficient background counts or contending with significant variance. In this work, we formalize this dilemma as a Rare Event Design (RED) problem: identifying optimal design parameters when the design metric to be minimized is inherently small. We then designed the Rare Event Surrogate Model (RESuM) for physics detector design optimization under RED conditions. RESuM uses a pretrained Conditional Neural Process (CNP) model to incorporate additional prior knowledges into a Multi-Fidelity Gaussian Process model. We applied RESuM to optimize neutron moderator designs for the LEGEND NLDBD experiment, identifying an optimal design that reduces neutron background by ($66.5\pm3.5$)\% while using only 3.3\% of the computational resources compared to traditional methods. Given the prevalence of RED problems in other fields of physical sciences, the RESuM algorithm has broad potential for simulation-intensive applications.
\end{abstract}

\section{Introduction}
Why is there more matter than antimatter in our universe? This question remains one of the most important yet unsolved questions in physics. Several Nobel Prizes have been awarded for groundbreaking discoveries that have advanced our understanding of this questions, including the discovery of CP violation in kaons \citep{nobel1980}, the detection of cosmic neutrinos \citep{nobel2002}, and the development of the Kobayashi-Maskawa theory of CP violation \citep{nobel2008}. Despite these monumental achievements, the reason for the dominance of matter over antimatter remains unsolved. One of the most promising next steps toward answering this question is the potential discovery of Neutrinoless Double-Beta Decay (NLDBD) \citep{dolinski20219}. Such a discovery would represent a major milestone in this direction and would undoubtedly be considered a Nobel-Prize-level breakthrough in physics. Due to its utmost importance, the entire U.S. nuclear physics community has gathered for a year-long discussion in 2023 and recommended the experimental search for NLDBD as the second-highest priority \citep{lrp} for next 10 years.

The most challenging aspect of NLDBD search is dealing with background events: physical events that are not NLDBD, but are indistinguishable from it. Since NLDBD is hypothesized to occur less than once every three years \citep{legend1000pCDR, dolinski20219}, even a single background event entering the detector could potentially ruin the entire detection effort. Therefore, designing ultra-pure NLDBD detectors with optimal parameters to minimize the probability of background events entering the detector becomes the utmost goal of all NLDBD experiments. Traditionally, the detector design procedure is conducted with simulations: we first simulate our detectors and $N_1$ background events under a certain design parameter $\theta_{1}$, then count the number of background events that eventually enter our detector, $m_1$. We then repeat the simulation process with another design parameter $\theta_{2}$ and count $m_2$. If $m_1/N_1 < m_2/N_2$, it suggests that the design $\theta_{1}$ is better than $\theta_{2}$. This simulation process can be repeated multiple times until an optimal design is found. An obvious shortcoming of this traditional approach is the computational cost: due to the ultra-pure nature of the NLDBD detector, $N$ needs to be extremely large ($\mathcal{O}(10^{4})$) for $m$ to even be above zero. This is amplified by the complexity of the design space, involving numerous and often non-linearly interdependent parameters such as detector geometry, material properties, and environmental conditions. 

An obvious solution to this problem is to build a surrogate model that can significantly accelerate our simulations \citep{mak2023, ravi2024}.  However, due to the rare event nature, $m$ is either 0 or a small, discrete integer, which leads to high variance in our design metric $m/N$. This variance renders training traditional continuous surrogate models extremely difficult. In this paper, we present our work in designing a new surrogate model that leverages the rare event design metric $m/N$ to navigate through a complex landscape and approximate the complex relationships between the design parameters~$\theta$ and $m/N$. This paper is structured as following: in Section~\ref{sec:related},  we discussed related work in physics surrogate model and CNP; in Section~\ref{section:RED_problem}, we formalize the aforementioned challenges as a Rare Event Design~(RED) problem using Poisson statistics; In Section~\ref{sec:resum}, we present our approach using the Conditional Neural Process (CNP) to incorporate additional prior information, leading to the development of the Rare Event Surrogate Model (RESuM). In Section~\ref{sec:exp}, we apply RESuM to a physics detector design problem in the LEGEND experiment—a world-leading  international experiment with 300 scientists in the search for NLDBD. The result shows that the RESuM model could reduce the LEGEND neutron background by $(66.5 \pm 3.5)$\% using only 3.3\% of the computational power compared to traditional methods. Lastly, in Section~\ref{sec:limitation_applications}, we discuss other possible domains, including Astronomy and Material Science, where the RESuM model could be applied due to the broad presence of RED problems in physical sciences.

\section{Related Works}\label{sec:related}
Due to the computational cost of particle physics simulations, generative models like VAE \citep{fu2024generative}, GAN \citep{duarte2021particlecloudgan, generative2018}, and diffusion models are widely used as surrogate models for fast simulation \citep{duarte2023evaluating}. Although these deep generative models, usually trained on large datasets, robustly reproduce enriched high-dimensional data, their black-box nature renders them non-interpretable and lacking clear statistical meaning. Meanwhile, the CNP model, as a probabilistic generative model, offers the distinct advantage of few-shot learning and provides clear statistical interpretation. It has demonstrated good performance in tackling few-shot problems, including classification tasks \citep{requeima2019cnpclassification}, statistical downscaling \citep{vaughan2022downscaling}, and hydrogeological modeling \citep{cui2022characterization}. In this study, we explore a novel surrogate modeling approach that focuses solely on key detector design metrics, leveraging a CNP model to extract meaningful insights from limited datasets.

\section{Rare Event Design Problem}\label{section:RED_problem}

\textbf{Definition}~Let \( \boldsymbol{\theta} \in \boldsymbol{\Theta} \) be the vector of design parameters, where \( \boldsymbol{\Theta} \) represents the space of all possible design parameters. Consider a simulation involving \( N\) events, or data points, under design parameter $\theta$; each event can either trigger a signal \footnote{``trigger a signal'' could represent any event of interest depending on the task setup. In the case of the NLDBD background minimization task, it means a background event successfully leach into the detector} or not.  Define a event-level random variable \( X\), where \( X_i = 1 \) if the $i$-th event triggers a signal and \( X_i = 0 \) if doesn't. 

Each simulated event $i$ is considered independent, and the outcome of \( X_i\) depends on two sets of parameters: a set of design parameters \( \theta \) which is universal across all events, and another sets of  event-specific parameters \( \boldsymbol{\phi_i} \in \boldsymbol{\Phi} \). The probability that the i-th event will trigger a signal is thereby defined as a function of both $\boldsymbol{\theta}$ and $\boldsymbol{\phi_i}$, which could be denoted as  $t (\boldsymbol{\theta}, \boldsymbol{\phi_i})$.

Let \( m \) represent the number of events that trigger a signal. The design metric \( y \) is then defined as:
\begin{equation}
    y = \frac{m}{N} = \frac{\sum_{i=1}^{N} X_i}{N}
\end{equation} 
\textbf{Rare Event Assumption}~The number of triggered events \( m \) follows a binomial distribution with the triggering probability \( t(\boldsymbol{\theta}, \boldsymbol{\phi_i}) \). Under the rare event assumption that \( m \ll N \) and the triggering probability for each event \( t(\boldsymbol{\theta}, \boldsymbol{\phi_i}) \) is small, the number of triggered events \( m \) can be approximated by a Poisson distribution as $m \sim \text{Poisson}\left(N\bar{t}(\boldsymbol{\theta})\right)$.  Where $\bar{t}(\theta)$ is the expected triggering probability over all simulated events when $N$ goes to infinity:
\begin{equation}
    \bar{t}(\boldsymbol{\theta}) = \int t(\boldsymbol{\theta}, \boldsymbol{\phi}) g(\boldsymbol{\phi}) d\boldsymbol{\phi}
\end{equation}
The function $g(\boldsymbol{\phi})$ denotes a predefined probability density function (PDF) where $\phi_i$ could be sampled from during the simulation process. $\bar{t}(\theta)$ is obtained by marginalizing $t(\theta,\phi)$ over $g(\phi)$. Therefore, the ultimate metric that we want to minimize is $\bar{t}$ , which is the expectation of $y$:
\begin{equation}
    \boldsymbol{\theta}^* = \arg \min_{\boldsymbol{\theta} \in \boldsymbol{\Theta}}\bar{t}(\boldsymbol{\theta})
\end{equation}
Since $\bar{t}$ depends on $\boldsymbol{\theta}$, minimizing  $\bar{t}$  requires extensive sampling of different $\boldsymbol{\theta}$ values within the design space \( \boldsymbol{\Theta} \) to identify the optimal parameter.

\textbf{Large N Scenario}~Assuming that $\bar{t}(\theta)$ remains fixed. When N becomes large, according to the central limit theorem, $m$ will tend to follow a normal distribution:
\[
m\sim \mathcal{N}(N\bar{t}(\boldsymbol{\theta}),N\bar{t}(\boldsymbol{\theta}))
\]
Since $y=m/N$, this means that $y$ will also follow a normal distribution with symmetric, well-defined statistical uncertainties $\bar{t}(\theta)/N$:
\[
y\sim \mathcal{N}(\bar{t}(\theta),\bar{t}(\theta)/N)
\]
As $N\xrightarrow[]{}\inf$, y will asymptotically approximate $\bar{t}(\theta)$ with statistical uncertainties approaching 0.

\textbf{Small N Scenario}~When \( N \) becomes small, the total number of instances \( m \) that trigger a signal has higher variance, as each individual instance has a significant impact on $m$. The accuracy measure \( y = \frac{m}{N} \) can no longer be approximated with a normal distribution. This makes $y$ more sensitive to statistical fluctuations of a few simulated events. Furthermore, there is a non-negligible probability that no event will trigger a signal, meaning that \( m = 0 \) and \(y \sim \frac{m}{N} = 0\). In summary, in the small N scenario, the design metric \( y \) of interests will only takes on a discrete set of values, \( y \in \left\{ \frac{0}{N}, \frac{1}{N}, \dots, \frac{m}{N} \right\} \). 

\section{Rare Event Surrogate Model}\label{sec:resum}
The Rare Event Surrogate Model~(RESuM) aims to solve the RED problem under the constraint of limited access to large \( N \) simulations. Consider a scenario where we run $K$ simulation trials with different design parameter $\theta$, indexed by \( k \); each simulation trial contains \( N \) events indexed by $i$.  The RESuM model includes three components: a Conditional Neural Process~(CNP) \citep{garnelo2018} model that is trained on event level; a Multi-Fidelity Gaussian Process~(MFGP) model that trains on simulation trial level; and active learning techniques to sequentially sample the parameter space after training. The conceptual framework and details of our model design are outlined in the following subsections.


\subsection{Bayesian Prior Knowledge with Conditional Neural Process}~\label{subsec:prior_knowledge}
The random variable \( X_{ki} \) represents whether the \( i^{\text{th}} \) event triggered a signal or not. In traditional particle physics, the value of \( X_{ki} \) is determined through a Monte Carlo simulation process: first, a parameter \(\phi_{ki}\) is sampled from the distribution \( g(\phi) \) to generate the event. This event then propagates through the detector, characterized by the design parameter \(\theta_k\). The outcome of the simulation, which implicitly involves the joint distribution \( t(\theta_k, \phi_{ki}) \), is only observed as \( X_{ki} \). As discussed before,  \( X_{ki} \) can only be  0 or 1. At small N scenario, the root cause of the discreteness of y is this binary nature: +1 if a signal is triggered or +0 if not. This produces significant statistical variance in $y$. 

Suppose we want to model this simulation process with a Bernoulli distribution:

\begin{equation}
X_{ki} \sim p(X_{ki}|\boldsymbol{\theta_k},\boldsymbol{\phi_{ki}}) = 
\text{Bernoulli}(p=t(\boldsymbol{\theta_k},\boldsymbol{\phi_{ki}}))
\label{eq:data-generation}
\end{equation}

The goal of incorporating prior knowledge is to turn the binary $X_{ki}$ into a continuous, floating-point score $\beta$ between 0 and 1. The score $\beta_{ki}$ should approximate $t(\theta_k,\phi_{ki})$ given design parameter \(\theta_k\) and event-specific parameter $\phi_{ki}$.




This work provides an alternative solution by adopting a similar idea to the CNP model.  CNP  works by learning a representation of input-output relationships from context data to predict outputs for new inputs \citep{garnelo2018}. In our case, the input is $\boldsymbol{\theta_k}$ and $\boldsymbol{\phi_{ki}}$, and the output is the random variable $X_{ki}$. The random process that generates $X_{ki}$ based on the inputs is the Bernoulli process controlled by \( t(\boldsymbol{\theta}, \boldsymbol{\phi}) \). We then adopt the same representation learning idea used in the CNP, which involves approximating the random process by sampling from a Gaussian distribution conditioned at observed data. The mean and variance are modeled with neural networks:
\begin{equation}
\text{Bernoulli}(p=t(\boldsymbol{\theta},\boldsymbol{\phi})) \approx \text{Bernoulli}(p=\beta)
\label{eq:nn-encoder-gaussian}
\end{equation}
\begin{equation}
\beta \sim \mathcal{N}(\mu_{NN}(\boldsymbol{\theta},\boldsymbol{\phi},\boldsymbol{w}), \sigma^2_{NN}(\boldsymbol{\theta},\boldsymbol{\phi},\boldsymbol{w}))|_{X_{ki},\boldsymbol{\phi_{ki}}, \boldsymbol{\theta_k}} 
\label{eq:beta_distribution}
\end{equation}
The nuisance parameters, denoted as $\boldsymbol{w}$, are optimized during the training of the neural networks by minimizing the likelihood of the observed data. Importantly, the neural networks are not trained to predict the binary observable $X$, but rather to estimate the continuous floating-point score $\beta$. A comprehensive description of the CNP model, along with the interpretation of the score $\beta$ and the associated loss function (likelihood), is provided in Appendix~\ref{subsec:link_vae}. The score $\beta$ for each simulated event serves as prior information that is incorporated into the Multi-Fidelity Gaussian Process (MFGP) surrogate model.

\subsection{Model Description}\label{subsec:resum_description}
\begin{figure}[hbt!]
    \centering
    \includegraphics[width=0.98\linewidth]{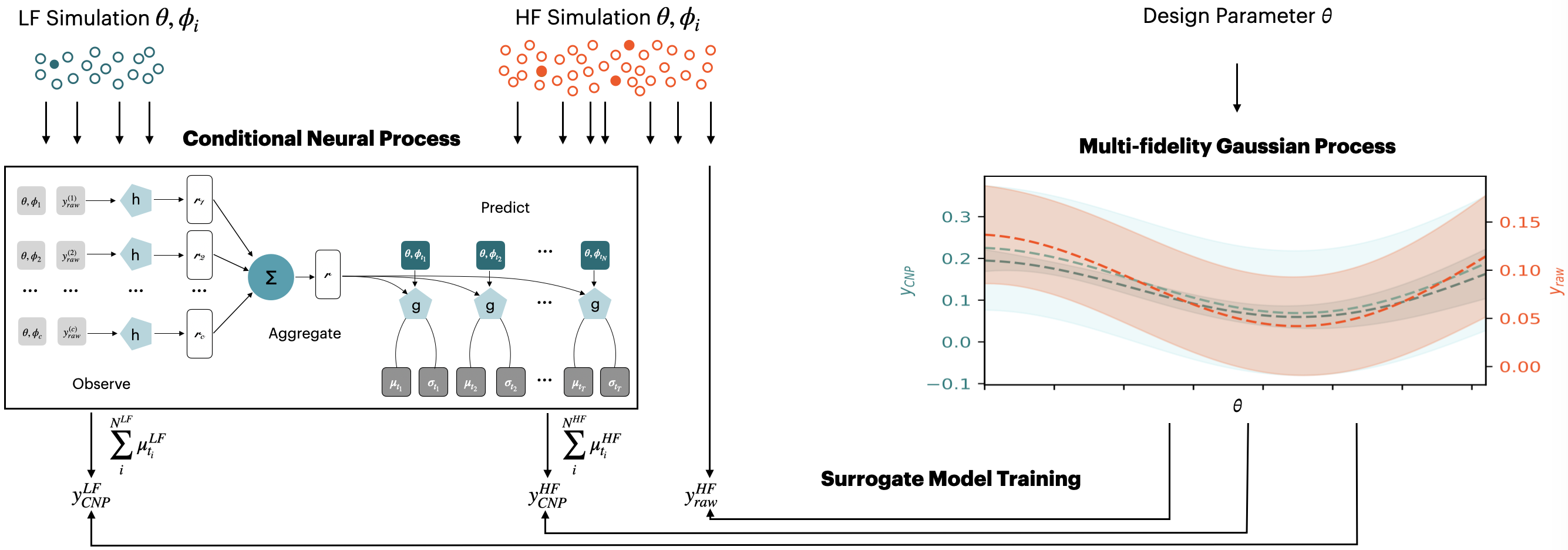}
    \caption{Overview of the RESuM framework for solving RED problems. The left side illustrates the CNP used for modeling both LF and HF simulation data. The CNP aggregates event-specific parameters \(\phi_i\) and design parameters \(\theta\) from LF and HF simulations to produce \(y_{CNP}^{LF}\) and \(y_{CNP}^{HF}\), which serve as inputs to the surrogate model. The right side shows the multi-fidelity Gaussian Process (MFGP) that combines predictions from LF and HF simulations to estimate the HF design metric \(y_{Raw}^{HF}\).}
    \label{fig:overview-cnp}
\end{figure}

Building on the conceptual framework described in~\ref{subsec:prior_knowledge}, we will provide an end-to-end overview of RESuM as shown in Figure~\ref{fig:overview-cnp}. We generate two types of simulations: low-fidelity (LF) and high-fidelity (HF). Detailed descriptions of these simulations can be found in Section~\ref{sec:MC-sim}. The primary distinction between them lies in the number of simulated events \( N \), where \( N_{HF} \gg N_{LF} \). Another key difference is the distribution \( g(\phi) \) from which the parameter \(\phi_i\) of each event is sampled, where HF simulation contains a more complicated, physics-oriented \( g(\phi) \). The low computational cost of LF simulation allows us to simulate more trials thereby exploring a broader range of $\theta$.

The first step is to train the CNP model. The CNP comprises three primary components: an encoder, an aggregator, and a decoder. The parameters $\theta_k,\phi_{ki}$, and $X_{ki}$ of each simulated event are first concatenated into a context point. The encoder, implemented as a Multi-Layer Perceptron (MLP), transforms each context point into a low-dimensional representation. These representations are then aggregated through averaging to form a unified representation that represents $t(\theta)$. The decoder uses $t(\theta)$ and the $\phi_{ki}$ of new data to output parameters \( \mu_{ki} \) and \( \sigma_{ki}^2 \) for each event \( i \). We then use \( \mu_{ki} \) and \( \sigma_{ki}^2 \) to form a normal distribution and sample a CNP score $\beta_{ki}$ from it. The scores $\beta_{ki}$ are chosen to naturally fit a normal-like distribution but bounded between 0 and 1. Since the CNP is trained at event level, $\beta_{ki}$ will be the same regardless of whether the event is generated in HF or LF simulation.


Based on the trained CNP model, the next step involves in calculating three design metrics at different fidelities. The first one is \(y_{Raw}=m/N\) from HF simulations, which is the ultimate design metric we want our surrogate model to emulate. The second metric is also derived from HF simulations but is defined as the average CNP score of all simulated events:
\begin{equation}
y_{CNP} = \frac{1}{N} \sum_{i=0}^{N} \beta_{ki}
\end{equation}
The third metric is $y_{CNP}$ calculated over LF simulations. These three design metrics are then incorporated into a Multi-Fidelity Gaussian Process~(MFGP) model \citep{kennedy2000,qian2008} to train the surrogate model. Co-kriging was used to account for correlations among different design metrics. The mathematical detail of MFGP can be found in Appendix~\ref{app:mfgp}

After training the MFGP model, we adopt active learning to select new sampling points \( \theta_{\text{new}} \) to generate $y_{Raw}$ with HF simulations. Since HF simulation is expensive, to determine which point to collect next, we use a gradient-based optimizer to find $\mathbf{ x}_{n+1} = \arg \max_{x\in\mathbb{X}} a(\mathbf{ x})$ \citep{emukit2023}. The acquisition function determines the next data point to explore by balancing exploration (high variance) and exploitation (high mean). We chose the integrated variance reduction method, where the next point, $\mathbf{x}_{n+1}$, is selected to maximally reduce the total variance of the model \citep{sacks1989}. More detail about the active learning method can be found in Appendix~\ref{subsec:active_learning}.
\section{Experiment and Result}\label{sec:exp}
\begin{figure}[h]
\begin{center}
    \centering
    \includegraphics[width=0.3\linewidth]{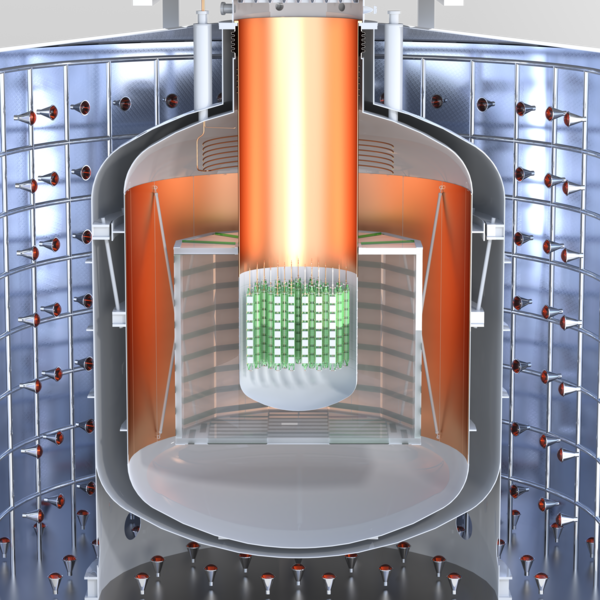}
    \includegraphics[width=0.3\linewidth]{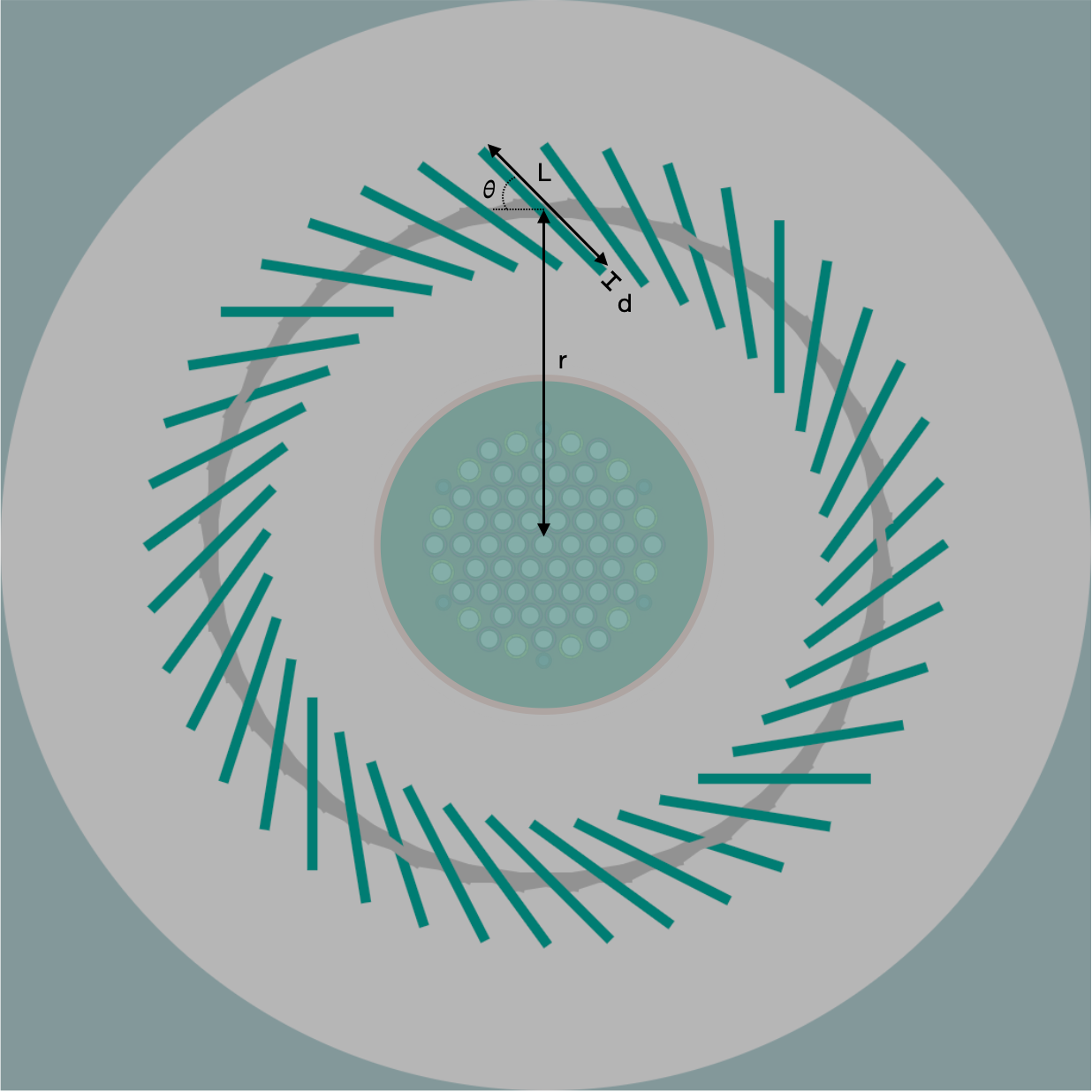}
    \caption{Left: Rendering of the \textsc{Legend-1000} experiment with the hundreds of detectors arranged in strings, operated in a bath of cryogenic ultra-pure liquid argon, showing a neutron moderator surrounding the inner cryostat which contains the detector strings \citep{legend-web}. Right: Illustration of a moderator design with the 5 design parameters (top view).}
    \label{fig:moderator-design}
\end{center}
\end{figure}

%
%
The Large Enriched Germanium Experiment for Neutrinoless Double-Beta Decay (\textsc{Legend}) is a next-generation, tonne-scale experiment using \( \rm ^{76}Ge \) detectors. LEGEND has been recognized by the nuclear physics community as one of the three leading international NLDBD experiments to be constructed in the next decade \citep{lrp}. In next 10 years, the experiment will be constructed by over 300 international collaborators. One of the primary background sources in LEGEND are \( \rm ^{77(m)}Ge \) \citep{legend1000pCDR, wiesinger2018}, which is created by neutrons entering the detector, having been produced by cosmic muons \citep{Pandola2007}. This background is particularly challenging, because it closely mimics NLDBD events, making it nearly impossible to distinguish and reject once produced. Currently, there are no efficient methods to eliminate this background category once it has been produced, aside from employing complex active tagging algorithms \citep{neuberger2021} that introduce additional dead time. The most viable solution to mitigate this background is through the design of a neutron moderator—a passive medium that reduces the incoming neutron flux before \( \rm ^{77(m)}Ge \) nuclei are produced. Figure~\ref{fig:moderator-design} provides an overview of the LEGEND detector and a proposed neutron moderator design. Using the presented model, our goal is to optimize the neutron moderator geometries, ensuring the polymer material effectively blocks the majority of neutrons, thereby preventing them from leaching into the detector. The optimization process consists of two key steps: generating simulations under different design parameters, and adopting the RESuM model to identify the optimal design of neutron moderators.

\subsubsection{Neutron Moderator Simulations}\label{sec:MC-sim}
As illustrated in Figure~\ref{fig:moderator-design} (Right), two geometric designs were proposed aiming at reducing neutron-induced backgrounds in the \textsc{Legend} detector: (1) a cylindrical layer of material surrounding the detector array, and (2) a turbine-like structure composed of panels. We define five design variables to include both geometries under continuous transition: the radius \texttt{r} of the cylindrical layer or the distance of the panels, the material thickness \texttt{d}, the number of panels \texttt{n}, the angle of the panels \(\varphi\), and the panel length \texttt{L}. These five parameters constitute the design parameter space \(\Theta\), where each parameter is allowed to vary within a predefined range. 

As discussed in Section~\ref{section:RED_problem}, a specific design \(\theta\) can be sampled from \(\Theta\) to perform simulations and obtain the corresponding design metric \(y_{Raw}\). For a given \(\theta\), we utilized a Monte Carlo (MC) simulation package based on the \textsc{GEANT-4} toolkit \citep{geant4_1,geant4_2}, integrated with the existing LEGEND software frameworks \citep{warwick-moritz, warwick-ramachers}. The MC simulation procedure is as follows: first, we modeled the entire LEGEND detector, including all its components, in GEANT4, as shown  in Figure~\ref{fig:moderator-design} (Left). Next, we generated a neutron moderator according to the sampled design parameter \(\theta\), which encapsulates the detectors, as shown in Figure~\ref{fig:moderator-design} (Right). In the subsequent step, \(N\) neutrons are simulated, each assigned an initial position \((x,y,z)\), momentum \((p_x,p_y,p_z)\), and energy $E$. These seven parameters form the event-specific parameter \(\phi\) and they are generated by sampling from a distribution $g(\phi)$, as discussed in Section~\ref{section:RED_problem}. Finally, the neutrons are allowed to propagate through the detector, and the number of \(\rm ^{77(m)}Ge\) nuclei produced, \(m\), is recorded. The design metric \(y_{Raw}\) is then computed as \(y_{Raw} = m / N\).

One major challenge is the high computational cost of the Monte Carlo (MC) simulations. A single run of the HF simulation requires approximately 170 CPU hours, making it impractical to fully explore the design space \(\Theta\) using a grid search approach. To address this, we implemented two levels of simulation fidelity: high-fidelity (HF) and low-fidelity (LF). The HF simulation provides our most accurate model of neutron generation and $\rm ^{77(m)}Ge$  production within LEGEND. In the HF simulation, the event-specific parameters $\phi_i$ of each neutron are sampled from a carefully-designed distribution $g(\phi)$, incorporating physical information. Since neutrons are primarily produced by cosmic muons descending from the atmosphere, the HF simulation starts by generating muons outside the \textsc{Legend} detector using site-specific muon flux and angular distributions provided by the \textsc{Musun} muon simulation software \citep{musun2009}. These muons, along with their secondary particle showers, propagate through the detector geometry, leading to the production of neutrons with their associated $\phi$. The total number of simulated neutrons \(N_{HF}\) for each design parameter \(\theta_k\) is typically very large, on the order of \(\mathcal{O}(10^7)\). 

The LF simulation, on the other hand, simplifies the complexity by skipping the muon simulation step. In this case, the parameters $\phi_i$ of each neutron are randomly sampled from a uniform distribution $g(\phi)$  in predefined range \footnote{To account for potential asymmetries, however, the neutron positions are adjusted based on predictions from the HF muon simulation, ensuring that the neutron placement remains as close as possible to realistic conditions}. The goal of the LF simulation is to estimate the production of $\rm ^{77(m)}Ge$ by tracking neutron propagation and background event production, without incorporating the additional complexity of muon physics in HF  $g(\phi)$. The total number of neutrons simulated, \(N_{LF}\), for each design parameter \(\theta\) in the LF simulation is on the order of \(\mathcal{O}(10^4)\), significantly smaller than in the HF simulation. 

While the LF simulation is based on a simplified uniform $g(\phi)$, it offers significant computational advantages, with a cost of only 0.15 CPU hours per run—about 1130 times faster than the HF simulation. This allows a broader exploration in the design parameter space \(\Theta\). Conversely, the HF simulations are crucial for providing rigorous estimates of the $\rm ^{77(m)}Ge$ background event production rate, ensuring that the optimized designs meet the stringent background requirements for the \textsc{Legend} experiment. In total, 4 HF and 304 LF simulations were generated to form the training dataset for surrogate model, with the four HF samples being reused from a previous simulation study, as suggested in \citet{neuberger2021}. 
\subsubsection{Conditional Neural Process Result}\label{sec:cnp}
The network structure and motivation of the CNP are discussed in Section~\ref{subsec:resum_description}. Training is performed using supervised learning, where a signal label (1) is assigned to neutrons that successfully produce \(\rm ^{77(m)}Ge\) background, and a background label (0) is assigned to neutrons that do not. A major challenge in training the CNP is the severe imbalance between signal and background, with a ratio of approximately \(1:5 \cdot 10^{4}\). This is consistent with the rare event assumption, where \(m \ll N\). To address this imbalance, we apply a data augmentation technique known as mixup \citep{zhang2018}, which helps create a more balanced and diverse dataset. Mixup generates new training samples \(\hat{x}\) by forming linear combinations of existing signal samples \(x_i\) and background samples \(x_j\), along with their corresponding labels \(y_i\) and \(y_j\):
\[
\hat{x} = \lambda x_i + (1 -\lambda)x_j \quad \text{and} \quad \hat{y} = \lambda y_i + (1-\lambda)y_j
\]
where \( \lambda \) is randomly drawn from a beta distribution \( B(0.1, 0.1) \). This process introduces a weighted blend of signal and background, helping to alleviate the imbalance in the data and improve the model’s generalization and robustness. 
%
%
\begin{figure}[hbt!]
    \centering
    \includegraphics[width=0.98\linewidth]{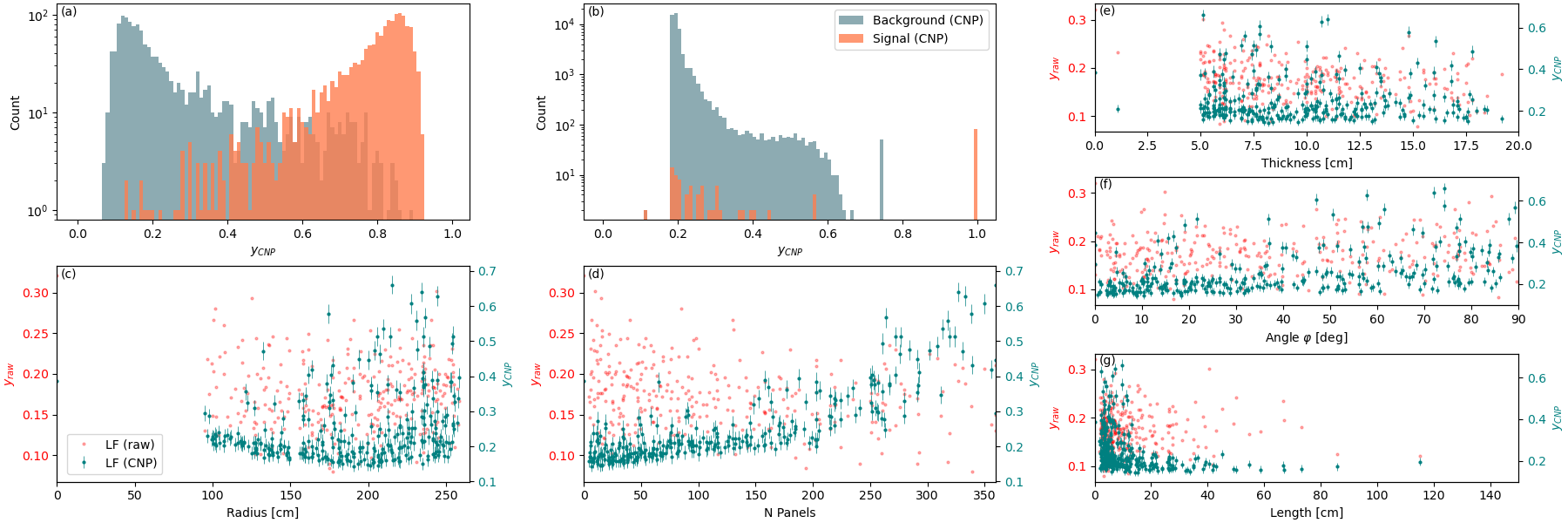}
    \caption{Comparison of raw data (red points) and CNP predictions (green points with error bars) across different design parameters. a) shows the results after 3540 iterations of CNP training, utilizing the described data augmentation method. b) demonstrates predicted scores for a validation dataset. c) to g) provides scatter plots of the raw metric \(y_{Raw}\) (red points) and \(y_{CNP}\) (green points with error bars), plotted against various design parameters: radius, number of panels, thickness, angle \(\varphi\), and length. The CNP model offers smoother, more refined predictions, effectively distinguishing signal from background across the different parameter spaces.}
    \label{fig:result-cnp}
\end{figure}
To demonstrate the effectiveness of the CNP in reducing statistical variance in the design metric, we calculated two different metrics for each LF simulation trial: the raw design metric \( y_{Raw} \) (in [$\rm nuclei/(kg \cdot yr)$]) and the averaged CNP score \( y_{CNP} \). The results of both metrics are plotted against five design parameters in Figure~\ref{fig:result-cnp}. As anticipated, the raw metric \( y_{Raw} \) exhibits significant statistical fluctuations that overshadows any correlations with respect to each design parameter. In contrast, the CNP score \( y_{CNP} \) reveals clear dependencies on the radius and number of panels (see Figure~\ref{fig:result-cnp} Bottom). This indicates that CNP effectively reveal additional prior information into the Multi-Fidelity Gaussian Process (MFGP) model.
\subsubsection{Surrogate Model Result}
\begin{figure}[hbt!]
    \begin{center}
    \includegraphics[width=0.98\linewidth]{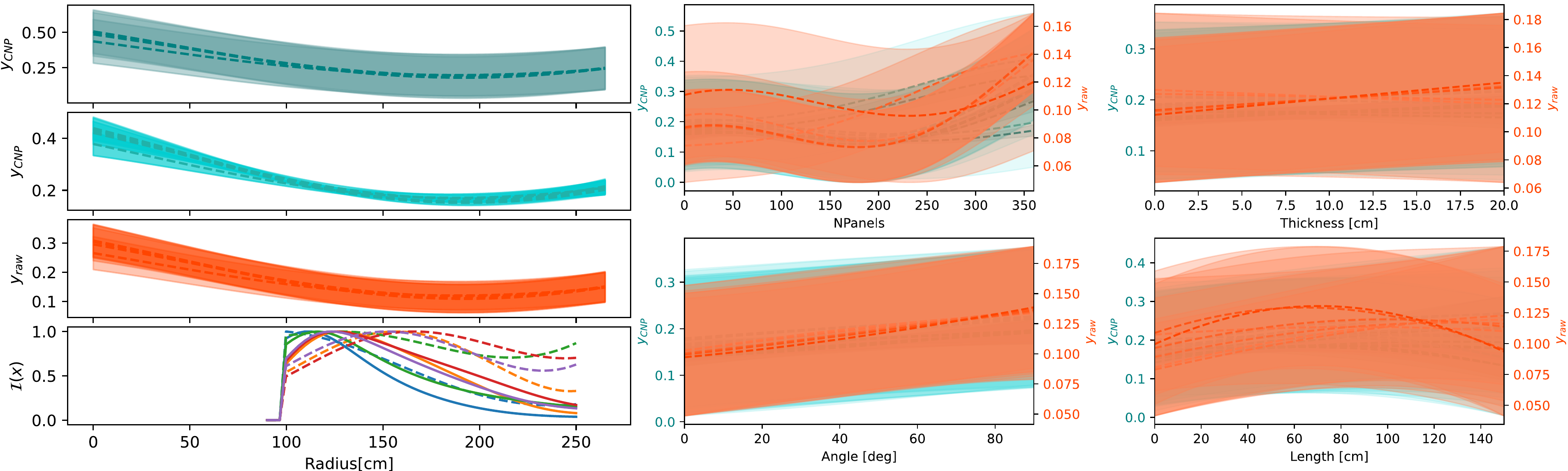}
    \caption{Left: One-dimensional CNP-LF (cyan), CNP-HF (dark cyan) and HF (orange) model predictions (dashed line) with uncertainty band (shaded area) as a function of the radius \texttt{r}. The lower panel shows the acquisition function as a function of the radius after each iteration. It guides the selection of future evaluation points in the input space to efficiently search for the optimal solution. Right: One dimensional model predictions as a function of the thickness, the panel's angle \texttt{$\varphi$}, the number of panels \texttt{n} and the length \texttt{L} at a fixed point in the design space. It illustrates the sequential model prediction update by adding new sampling points in each iteration.}
    \label{fig:model-update}
    \end{center}
\end{figure}

The surrogate model was trained with three design metrics at different fidelities: $y_{Raw}$ for HF simulations, $y_{CNP}$ for HF simulations, and $y_{CNP}$ for LF simulations. The ultimate goal is to emulate $y_{Raw}$ for HF simulations which provides a more accurate representation of $\rm ^{77(m)}Ge$ production rates under design parameter $\theta$.  The Multi- Fidelity Gaussian Process model was carried out by using the Emukit python library \citep{emukit2023} which offers a high level interface for solving optimization problems.  
Figure~\ref{fig:model-update} (Left Bottom) illustrates the active learning process using an acquisition function. It displays the HF model prediction (Top) and the acquisition function (Bottom) as a function of the radius after each iteration. The surrogate model provides an estimate of the design metric of interest ($y_{Raw}$ from HF simulations) along with its associated uncertainty (shown as shaded area) at each point in the input space. The acquisition function from active learning evolves as more data points are added, progressively refining the objective function approximations by calculating the difference between the best observed value and the surrogate model's prediction, while also accounting for the uncertainty in its predictions. As notable in the lower panel of Figure~\ref{fig:model-update} (Left), the acquisition function initially explores regions with high uncertainty, especially at medium distances where the optimum is likely to be found. The active learning procedure can be iterated as long as computation resources are available. In this work, six active learning iterations were performed to obtain the final result. 
The HF model predictions for the \( \rm ^{77(m)}Ge \) production rate are shown in Figure~\ref{fig:model-update}, which displays one-dimensional projections of $y_{CNP}$ from LF simulations (cyan),  $y_{CNP}$ from HF simulations (dark cyan), and  $y_{Raw}$ from  HF simulations (orange) model predictions (dashed line) along with associated uncertainties (shaded area), as functions of radius \( r \), thickness \( d \), panel angle \( \varphi \), number of panels \( n \), and length \( L \), presented in reading order from left-to-right, top-to-bottom. These figures illustrate how the model’s uncertainty decreased with each new sampling iteration, particularly where  significant improvements in model certainty were observed. After six active learning iterations, the model predictions converged on several optimal designs, as shown in Table~\ref{tab:result}. These designs exhibit a range of configurations, with most favoring smaller panel angles \(\varphi\) and a higher number of shorter panels, while one achieve optimal performance with fewer but significantly longer panels. Additionally, the optimal designs tend to cluster around two radii ranges, approximately 165 cm and 200 cm, suggesting that positioning the neutron moderator near these distances provides the best balance between effective neutron capture and material efficiency. This positioning allows for sufficient moderator mass to trap neutrons while maintaining an appropriate distance to minimize neutron escape. These designs outperformed those with larger gaps between panels, primarily due to the increased mass of neutron moderator material, which enhances neutron capture and minimizes background radiation. The optimal design reduces the \( \rm ^{77(m)}Ge \) production rate from 0.238 \( \rm nuclei/(kg \cdot yr) \) to 0.0798 \( \rm nuclei/(kg \cdot yr) \), leading to a $(66.5 \pm 3.5)$\% reduction in neutron-induced background in LEGEND.
%
\begin{table}[hbt!]
\caption{Optimal neutron moderator design parameters identified by the RESuM model for the LEGEND experiment. The table displays the key design parameters, including radius \texttt{r}, panel thickness \texttt{d}, number of panels \texttt{n}, panel angle \(\varphi\), and length \texttt{L}. These parameters were determined based on RESuM’s ability to optimize neutron background reduction.}
\label{tab:result}
\begin{center}
\footnotesize
\begin{tabular}{|c|c|c|c|c|c|c|} \hline
\texttt{r} [cm] & \texttt{d} [cm] & n & $\varphi$ [deg] & \texttt{L} [cm]  & $y^{min}_{raw}$ $[ \rm nuclei/(kg \cdot yr) ]$ & $\sigma^{min}_{raw}$ $[ \rm nuclei/(kg \cdot yr) ]$ \\ \hline
165.6 & 3.3 & 188 & 19.3 & 7.5   & 0.0798  & 0.0483 \\ \hline
207.3 & 2.8 & 120 & 3.5  & 3.2   & 0.0786  & 0.0494 \\ \hline
202.2 & 2.4 & 153 & 9.1  & 3.0   & 0.0787  & 0.0489 \\ \hline
164.3 & 4.2 & 192 & 15.4 & 3.1   & 0.0784 & 0.0485 \\ \hline
198.3 & 3.2 & 11  & 8.0  & 106.5 & 0.0779 & 0.0483 \\ \hline
145.5 & 3.4 & 193 & 17.9 & 1.7 & 0.0809 & 0.0489 \\ \hline
\end{tabular}
\end{center}
\end{table}
%
 %
%
Notably, RESuM identified the optimal design parameters with drastically reduced computational costs. Each HF simulation required 170 CPU hours, while each LF simulation needed just 0.15 CPU hour. If we were to explore the design space with only HF simulations, traditional methods would have required 52,700 CPU hours to explore all 310 design parameter sets. In contrast, RESuM used 310 LF simulations and 10 HF (4 for MFGP training and 6 for active learning) simulations, totaling 1746.5 CPU hours—only 3.3\% of the computational resources required by conventional approaches. 

\subsubsection{Surrogate Model Validation}
The ultimate goal of RESuM is to emulate the $y_{Raw}$ values of HF simulations given the input of design parameters $\theta$ at any location within the design space $\Theta$. To ensure the accuracy of the learned mapping between $\theta$ and $y_{Raw}$, additional independent, out-of-sample HF simulations are required for validation. Given the high computational demands of HF simulations and our limited resources, we generated 100 out-of-sample HF simulations at randomly sampled $\theta$ values. The $y_{Raw}$ for each validation simulation is determined as the ground truth by counting the number of $\rm ^{77(m)}Ge$ nuclei produced. Simultaneously, the $\theta$ value from each validation simulation was input into the trained RESuM model to predict $\hat{y}_{Raw}$ along with the associated uncertainty $\hat{\sigma}_{Raw}$ generated by the Gaussian Process. The validation study aims to assess whether $\hat{y}_{Raw} \pm \hat{\sigma}_{Raw}$ accurately covers the ground truth $y_{Raw}$ across 1, 2, and 3 $\hat{\sigma}_{Raw}$.

\begin{figure}[hbt!]
    \centering
    \includegraphics[width=0.98\linewidth]{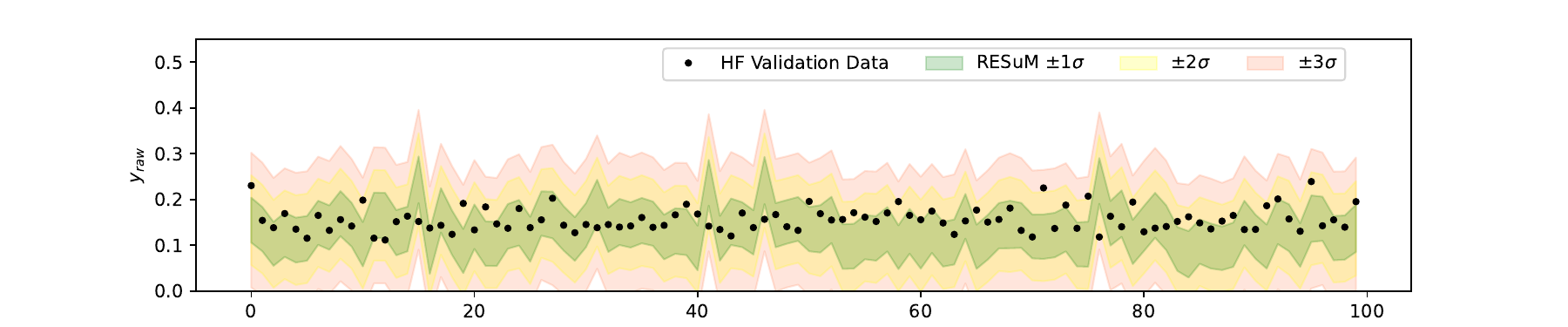}
    \caption{Comparison of HF simulation validation data (black dots) with the RESuM model predictions. The shaded regions represent the uncertainty bands at different confidence levels: $\pm 1\hat{\sigma}$ (green), $\pm 2\hat{\sigma}$ (yellow), and $\pm 3\hat{\sigma}$ (red). The RESuM model captures the overall trend of the HF validation data, with 69\% of the points falling within the  $\pm 1\sigma$  confidence band, indicating good agreement between the model predictions and validation data.}
    \label{fig:validation}
\end{figure}
The coverage results are shown in Figure~\ref{fig:validation}. The plot indicates that 69\% of the ground truth \( y_{Raw} \) falls within the 1 \(\sigma_{Raw}\) band of \(\hat{y_{Raw}}\), 95\% falls within the 2 \(\sigma_{Raw}\) band, and 100\% falls within the 3 \(\sigma_{Raw}\) band. Assuming a standard normal distribution, the expected coverage at 1, 2, and 3 \(\sigma\) is 68.27
\%, 95.45\%, and 99.73\% , respectively. These results indicate that the RESuM model achieves proper coverage at 1,2, and 3 \(\hat{\sigma}_{Raw}\). For comparison, we conducted a study without $y_{CNP}$ in the RESuM model: we trained an MFGP surrogate model solely on $y_{Raw}$ from the LF and HF simulations. This resulted in significantly poor coverage, as detailed in Appendix~\ref{subsec:mf_comparsion}.
This further demonstrates that the overall agreement between the ground truth and RESuM predictions remains good, highlighting the RESuM model's capability to surrogate complex detector design simulations.

\section{Limitations and applications}\label{sec:limitation_applications}
\paragraph{Limitations:} The primary limitation of this work is the restricted computational resources available for thoroughly validating the model’s performance. Due to these constraints, we were limited to generating only 100 HF simulations to assess coverage. Ideally, with unlimited computational resources, a comprehensive grid search across the 5-dimensional design parameter space would allow for more robust validation. Additionally, the active learning strategy employed in RESuM is relatively simplistic. Future work will focus on exploring more sophisticated active learning algorithms to more efficiently identify the optimal design.

\paragraph{Generalizability:} 
Although this work focuses on a specific detector design scenario within the LEGEND experiment, we believe that the mathematical formulation of RED problem, as outlined in Section~\ref{section:RED_problem}, is applicable to a wide range of simulation and optimization challenges in the physical sciences. A few examples are provided below:
\begin{itemize}
    \item \textbf{Astromony:} In computational astronomy, an emerging area involves simulating binary black hole (BBH) mergers to match the gravitational wave (GW) signals detected by the LIGO experiments \citep{fishbach2017}. This process involves highly complex and computationally expensive many-body simulations \citep{kruckow20218}. In this context, cosmological constants, such as the Hubble Constant and Dark Energy Density, can be treated as design parameters \( \theta \), while the initial position, mass, and spin of each black hole are considered event-specific parameters \( \phi \). The design metric \( m \) is defined as the number of BBH mergers occuring over a given time period, with N representing the black holes simulated. The ground truth is provided by the GW catalog from LIGO. Some related work in this direction include \citep{lin2021, broekgaarden2019}.
    \item \textbf{Material Science}: First-principles simulations using Density Functional Theory (DFT) are widely employed in condensed matter research to predict novel material properties and optimize designs \citep{dovesi2018quantum,atomicsimulation2014, kang2019firstprincipledesign}. These computationally expensive simulations depend on various parameters such as temperature, pressure, and doping concentration. Despite the complicated simulation, the metrics derived from DFT calculations typically consist of a few quantities, such as electronic band structures. The computational intensity of DFT often limits extensive parameter space exploration, particularly for complex materials or large-scale systems. This constraint necessitates the development of efficient surrogate models to rapidly approximate DFT results while maintaining acceptable accuracy, potentially accelerating materials discovery and optimization processes.

\end{itemize}

\section{Conclusion and Outlook}
In this work, we presented RESuM, a rare event surrogate model designed for detector design optimization problems in physics. We began by statistically define the RED problem and proposed a CNP-enhanced surrogate model to solve it. We demonstrated the effectiveness of RESuM on a specific task: optimizing the neutron moderator design for the LEGEND NLDBD experiment. Our results show that RESuM successfully identified an optimal design, reducing neutron background by 66.5\% while utilizing only 3.3\% of the computational resources required by traditional methods. The accuracy and coverage of the trained RESuM model were successfully validated with independently simulated HF datasets. This means the surrogate model aligns well with physical simulations with proper coverage. Based on the statistical formulation and validation results, we believe that the RESuM model is more statistically robust and interpretable compared to other surrogate models, such as those based on GANs, for accelerating simulations \citep{oliveira2017,fu2024generative}.

Although this work focuses on a specific detector design problem in physics, we believe that RED problems are prevalent in many other domains, as discussed in Section~\ref{sec:limitation_applications}. Our future work will focus on two key directions: first, we want to further refine the RESuM model by validating with more HF simulations and improving the active learning algorithm; second, we intend to explore additional application of the RESuM model, especially in simulating Binary Black Hole mergers in Astronomy, as outlined in Section~\ref{sec:limitation_applications}. This effort seeks to foster greater collaboration across the machine learning, physics, and astronomy communities, ultimately benefiting all fields involved.

\section{Reproducibility Statement}
To ensure that the RESuM model can be reliably reproduced, we have carefully documented all aspects of the methodology and experiment. The simulation tool and package used in our work is explicitly referenced in Section~\ref{sec:MC-sim}. The dataset, preprocessing steps, and model architecture are described in Section~\ref{subsec:resum_description}, Section~\ref{sec:MC-sim} and Section~\ref{sec:cnp}. Model parameters and evaluation metrics are clearly defined in Section \ref{subsec:resum_description} and Section~\ref{sec:cnp}. The code of this work is anonymized and released as the supplementary material of this submission. All scripts for data handling, model training, and evaluation are included in the supplementary material, along with environment specifications and fixed random seeds to minimize variability. The training data of this work is too large as it involves in expensive simulations. The authors plan to release training data in the camera-ready version. 

\bibliography{resum_references}

\begin{thebibliography}{39}
\providecommand{\natexlab}[1]{#1}
\providecommand{\url}[1]{\texttt{#1}}
\expandafter\ifx\csname urlstyle\endcsname\relax
  \providecommand{\doi}[1]{doi: #1}\else
  \providecommand{\doi}{doi: \begingroup \urlstyle{rm}\Url}\fi

\bibitem[Agostinelli et~al.(2003)]{geant4_1}
S.~Agostinelli et~al.
\newblock {GEANT4--a simulation toolkit}.
\newblock \emph{Nucl. Instrum. Meth. A}, 506:\penalty0 250--303, 2003.
\newblock \doi{10.1016/S0168-9002(03)01368-8}.

\bibitem[Allison et~al.(2006)]{geant4_2}
J.~Allison et~al.
\newblock Geant4 developments and applications.
\newblock \emph{IEEE Transactions on Nuclear Science}, 53\penalty0 (1):\penalty0 270--278, 2006.
\newblock \doi{10.1109/TNS.2006.869826}.

\bibitem[Broekgaarden et~al.(2019)Broekgaarden, Justham, de~Mink, Gair, Mandel, Stevenson, Barrett, Vigna-Gómez, and Neijssel]{broekgaarden2019}
F.~S. Broekgaarden, S.~Justham, S.~E. de~Mink, J.~Gair, I.~Mandel, S.~Stevenson, J.~W. Barrett, A.~Vigna-Gómez, and C.~J. Neijssel.
\newblock {stroopwafel: simulating rare outcomes from astrophysical populations, with application to gravitational-wave sources}.
\newblock \emph{Monthly Notices of the Royal Astronomical Society}, 490\penalty0 (4):\penalty0 5228--5248, 09 2019.
\newblock ISSN 0035-8711.
\newblock \doi{10.1093/mnras/stz2558}.
\newblock URL \url{https://doi.org/10.1093/mnras/stz2558}.

\bibitem[Committee(2023)]{lrp}
Nuclear Science~Advisory Committee.
\newblock A new era of discovery: The 2023 long range plan for nuclear science, 2023.
\newblock URL \url{https://nuclearsciencefuture.org/wp-content/uploads/2024/02/23-G06476-2024-LRP-17x11-pcg-1.24.24.pdf}.

\bibitem[Cronin and Fitch(1980)]{nobel1980}
J.~Cronin and V.~Fitch.
\newblock The nobel prize in physics 1980: Discovery of cp violation in the decay of neutral kaons, 1980.
\newblock URL \url{https://www.nobelprize.org/prizes/physics/1980/summary/}.
\newblock Accessed: 2024-09-11.

\bibitem[Cui et~al.(2022)Cui, Chen, and Liu]{cui2022characterization}
Z.~Cui, Q.~Chen, and G.~Liu.
\newblock Characterization of subsurface hydrogeological structures with convolutional conditional neural processes on limited training data.
\newblock \emph{Water Resources Research}, 58\penalty0 (12):\penalty0 e2022WR033161, 2022.

\bibitem[de~Oliveira et~al.(2017)de~Oliveira, Paganini, and Nachman]{oliveira2017}
L.~de~Oliveira, M.~Paganini, and B.~Nachman.
\newblock Learning particle physics by example: Location-aware generative adversarial networks for physics synthesis.
\newblock \emph{Computing and Software for Big Science}, 1, 09 2017.
\newblock \doi{10.1007/s41781-017-0004-6}.

\bibitem[Dolinski et~al.(2019)Dolinski, Poon, and Rodejohann]{dolinski20219}
M.~J. Dolinski, A.W.P. Poon, and W.~Rodejohann.
\newblock Neutrinoless double-beta decay: Status and prospects.
\newblock \emph{Annual Review of Nuclear and Particle Science}, 69\penalty0 (Volume 69, 2019):\penalty0 219--251, 2019.
\newblock ISSN 1545-4134.
\newblock \doi{https://doi.org/10.1146/annurev-nucl-101918-023407}.
\newblock URL \url{https://www.annualreviews.org/content/journals/10.1146/annurev-nucl-101918-023407}.

\bibitem[Dovesi et~al.(2018)Dovesi, Erba, Orlando, Zicovich-Wilson, Civalleri, Maschio, R{\'e}rat, Casassa, Baima, Salustro, et~al.]{dovesi2018quantum}
R.~Dovesi, A.~Erba, R.~Orlando, C.~M. Zicovich-Wilson, B.~Civalleri, L.~Maschio, M.~R{\'e}rat, S.~Casassa, J.~Baima, S.~Salustro, et~al.
\newblock Quantum-mechanical condensed matter simulations with crystal.
\newblock \emph{Wiley Interdisciplinary Reviews: Computational Molecular Science}, 8\penalty0 (4):\penalty0 e1360, 2018.

\bibitem[Fishbach and Holz(2017)]{fishbach2017}
M.~Fishbach and D.~E. Holz.
\newblock Where are ligo’s big black holes?
\newblock \emph{The Astrophysical Journal Letters}, 851\penalty0 (2):\penalty0 L25, dec 2017.
\newblock \doi{10.3847/2041-8213/aa9bf6}.
\newblock URL \url{https://dx.doi.org/10.3847/2041-8213/aa9bf6}.

\bibitem[Garnelo et~al.(2018)Garnelo, Rosenbaum, Maddison, Ramalho, Saxton, Shanahan, Teh, Rezende, and Eslami]{garnelo2018}
M.~Garnelo, D.~Rosenbaum, C.~J. Maddison, T.~Ramalho, D.~Saxton, M.~Shanahan, Y.~Whye Teh, D.~J. Rezende, and S.~M.~Ali Eslami.
\newblock Conditional neural processes, 2018.
\newblock URL \url{https://arxiv.org/abs/1807.01613}.

\bibitem[Hutter et~al.(2014)Hutter, Iannuzzi, Schiffmann, and VandeVondele]{atomicsimulation2014}
J.~Hutter, M.~Iannuzzi, F.~Schiffmann, and J.~VandeVondele.
\newblock cp2k: atomistic simulations of condensed matter systems.
\newblock \emph{Wiley Interdisciplinary Reviews: Computational Molecular Science}, 4\penalty0 (1):\penalty0 15--25, 2014.

\bibitem[Kang et~al.(2019)Kang, Liang, Jiang, Lin, and Chen]{kang2019firstprincipledesign}
L.~Kang, F.~Liang, X.~Jiang, Z.~Lin, and C.~Chen.
\newblock First-principles design and simulations promote the development of nonlinear optical crystals.
\newblock \emph{Accounts of Chemical Research}, 53\penalty0 (1):\penalty0 209--217, 2019.

\bibitem[Kansal et~al.(2021)Kansal, Duarte, Su, Orzari, Tomei, Pierini, Touranakou, Gunopulos, et~al.]{duarte2021particlecloudgan}
R.~Kansal, J.~Duarte, H.~Su, B.~Orzari, T.~Tomei, M.~Pierini, M.~Touranakou, D.~Gunopulos, et~al.
\newblock Particle cloud generation with message passing generative adversarial networks.
\newblock \emph{Advances in Neural Information Processing Systems}, 34:\penalty0 23858--23871, 2021.

\bibitem[Kansal et~al.(2023)Kansal, Li, Duarte, Chernyavskaya, Pierini, Orzari, and Tomei]{duarte2023evaluating}
R.~Kansal, A.~Li, J.~Duarte, N.~Chernyavskaya, M.~Pierini, B.~Orzari, and T.~Tomei.
\newblock Evaluating generative models in high energy physics.
\newblock \emph{Physical Review D}, 107\penalty0 (7):\penalty0 076017, 2023.

\bibitem[Kennedy and O’Hagan(2000)]{kennedy2000}
M.~C. Kennedy and A.~O’Hagan.
\newblock Predicting the output from a complex computer code when fast approximations are available.
\newblock \emph{Biometrika}, vol. 87\penalty0 (no. 1):\penalty0 1–13, 2000.

\bibitem[Kingma(2013)]{vae}
D.~P. Kingma.
\newblock Auto-encoding variational bayes.
\newblock \emph{arXiv preprint arXiv:1312.6114}, 2013.

\bibitem[Kobayashi and Maskawa(2008)]{nobel2008}
M.~Kobayashi and T.~Maskawa.
\newblock The nobel prize in physics 2008: Discovery of the origin of cp violation, 2008.
\newblock URL \url{https://www.nobelprize.org/prizes/physics/2008/summary/}.
\newblock Accessed: 2024-09-11.

\bibitem[Koshiba(2002)]{nobel2002}
M.~Koshiba.
\newblock The nobel prize in physics 2002: Detection of cosmic neutrinos, 2002.
\newblock URL \url{https://www.nobelprize.org/prizes/physics/2002/summary/}.
\newblock Accessed: 2024-09-11.

\bibitem[Kruckow et~al.(2018)Kruckow, Tauris, Langer, Kramer, and Izzard]{kruckow20218}
M.~U. Kruckow, T.~M. Tauris, N.~Langer, M.~Kramer, and R.~G. Izzard.
\newblock {Progenitors of gravitational wave mergers: binary evolution with the stellar grid-based code ComBinE}.
\newblock \emph{Monthly Notices of the Royal Astronomical Society}, 481\penalty0 (2):\penalty0 1908--1949, 08 2018.
\newblock ISSN 0035-8711.
\newblock \doi{10.1093/mnras/sty2190}.
\newblock URL \url{https://doi.org/10.1093/mnras/sty2190}.

\bibitem[Kudryavtsev(2009)]{musun2009}
V.A. Kudryavtsev.
\newblock Muon simulation codes music and musun for underground physics.
\newblock \emph{Comput. Phys. Commun.}, 180\penalty0 (3):\penalty0 339--346, 2009.
\newblock ISSN 0010-4655.
\newblock \doi{https://doi.org/10.1016/j.cpc.2008.10.013}.

\bibitem[LEGEND-Collaboration et~al.()]{legend-web}
LEGEND-Collaboration et~al.
\newblock URL \url{https://legend-exp.org/science/legend-pathway/legend-1000}.

\bibitem[LEGEND-Collaboration et~al.(2021)]{legend1000pCDR}
LEGEND-Collaboration et~al.
\newblock Legend-1000 preconceptual design report, 2021.
\newblock URL \url{https://arxiv.org/abs/2107.11462}.

\bibitem[Li et~al.(2023)Li, Mak, Paquet, and Bass]{mak2023}
K.~Li, S.~Mak, J.~Paquet, and S.~Bass.
\newblock Additive multi-index gaussian process modeling, with application to multi-physics surrogate modeling of the quark-gluon plasma, 06 2023.

\bibitem[Lin et~al.(2021)Lin, Bingham, Broekgaarden, and Mandel]{lin2021}
L.~Lin, D.~Bingham, F.~Broekgaarden, and I.~Mandel.
\newblock Uncertainty quantification of a computer model for binary black hole formation, 06 2021.

\bibitem[Neuberger()]{warwick-moritz}
M.~Neuberger.
\newblock warwick-legend.
\newblock \url{https://github.com/MoritzNeuberger/warwick-legend}.

\bibitem[Neuberger et~al.(2021)]{neuberger2021}
M.~Neuberger et~al.
\newblock \emph{J. Phys.: Conf. Ser.}, 2156\penalty0 (012216), 2021.

\bibitem[Paleyes et~al.(2023)Paleyes, Mahsereci, and Lawrence]{emukit2023}
A.~Paleyes, M.~Mahsereci, and N.~D. Lawrence.
\newblock Emukit: A {P}ython toolkit for decision making under uncertainty.
\newblock \emph{Proc. Python Sci. Conf.}, 2023.

\bibitem[Pandola et~al.(2007)Pandola, Bauer, Kröninger, Liu, Tomei, Belogurov, Franco, Klimenko, and Knapp]{Pandola2007}
L.~Pandola, M.~Bauer, K.~Kröninger, X.~Liu, C.~Tomei, S.~Belogurov, D.~Franco, A.~Klimenko, and M.~Knapp.
\newblock Monte carlo evaluation of the muon-induced background in the gerda double beta decay experiment.
\newblock \emph{Nucl. Instrum. Meth. A}, 570\penalty0 (1):\penalty0 149--158, 2007.
\newblock ISSN 0168-9002.
\newblock \doi{https://doi.org/10.1016/j.nima.2006.10.103}.
\newblock URL \url{https://www.sciencedirect.com/science/article/pii/S0168900206018031}.

\bibitem[Qian and Wu(2008)]{qian2008}
P.Z.G. Qian and C.F.J. Wu.
\newblock Bayesian hierarchical modeling for integrating low-accuracy and high-accuracy experiments.
\newblock \emph{Technometrics}, 50\penalty0 (2):\penalty0 192--204, 2008.
\newblock URL \url{https://doi.org/10.1198/004017008000000082}.

\bibitem[Ramachers and Morgan()]{warwick-ramachers}
Y.~Ramachers and B.~Morgan.
\newblock warwick-legend.
\newblock \url{https://github.com/drbenmorgan/warwick-legend}.

\bibitem[Ravi et~al.(2024)Ravi, Fediukov, Dietrich, Neckel, Buse, Bergmann, and Bungartz]{ravi2024}
K.~Ravi, V.~Fediukov, F.~Dietrich, T.~Neckel, F.~Buse, M.~Bergmann, and H.-J. Bungartz.
\newblock Multi-fidelity gaussian process surrogate modeling for regression problems in physics, 2024.
\newblock URL \url{https://arxiv.org/abs/2404.11965}.

\bibitem[Requeima et~al.(2019)Requeima, Gordon, Bronskill, Nowozin, and Turner]{requeima2019cnpclassification}
J.~Requeima, J.~Gordon, J.~Bronskill, S.~Nowozin, and R.E. Turner.
\newblock Fast and flexible multi-task classification using conditional neural adaptive processes.
\newblock \emph{Advances in neural information processing systems}, 32, 2019.

\bibitem[Sacks et~al.(1989)Sacks, Welch, Mitchell, and Wynn]{sacks1989}
J.~Sacks, W.~J. Welch, T.~J. Mitchell, and H.~P. Wynn.
\newblock {Design and Analysis of Computer Experiments}.
\newblock \emph{Statistical Science}, 4\penalty0 (4):\penalty0 409 -- 423, 1989.
\newblock \doi{10.1214/ss/1177012413}.
\newblock URL \url{https://doi.org/10.1214/ss/1177012413}.

\bibitem[Vallecorsa(2018)]{generative2018}
S.~Vallecorsa.
\newblock Generative models for fast simulation.
\newblock In \emph{Journal of Physics: Conference Series}, volume 1085, page 022005. IOP Publishing, 2018.

\bibitem[Vaughan et~al.(2022)Vaughan, Tebbutt, Hosking, and Turner]{vaughan2022downscaling}
A.~Vaughan, W.~Tebbutt, J.~S. Hosking, and R.~E. Turner.
\newblock Convolutional conditional neural processes for local climate downscaling.
\newblock \emph{Geoscientific Model Development}, 15\penalty0 (1):\penalty0 251--268, 2022.

\bibitem[Wiesinger et~al.(2018)Wiesinger, Pandola, and Schönert]{wiesinger2018}
Ch. Wiesinger, L.~Pandola, and S.~Schönert.
\newblock Virtual depth by active background suppression: revisiting the cosmic muon induced background of gerda phase~{II}.
\newblock \emph{Eur. Phys. J. C}, 78\penalty0 (7), 2018.
\newblock \doi{10.1140/epjc/s10052-018-6079-3}.
\newblock URL \url{https://doi.org/10.1140%2Fepjc%2Fs10052-018-6079-3}.

\bibitem[{Z. Fu} et~al.(2024){Z. Fu}, {Ch. Grant}, {D. M. Krawiec}, {A. Li}, and {L. A. Winslow}]{fu2024generative}
{Z. Fu}, {Ch. Grant}, {D. M. Krawiec}, {A. Li}, and {L. A. Winslow}.
\newblock Generative models for simulation of kamland-zen.
\newblock \emph{The European Physical Journal C}, 84\penalty0 (6):\penalty0 651, 2024.

\bibitem[Zhang et~al.(2018)Zhang, Cisse, Dauphin, and Lopez-Paz]{zhang2018}
H.~Zhang, M.~Cisse, Y.~N. Dauphin, and D.~Lopez-Paz.
\newblock mixup: Beyond empirical risk minimization, 2018.
\newblock URL \url{https://arxiv.org/abs/1710.09412}.

\end{thebibliography}
\bibliographystyle{plainnat}

\section*{Appendix}
\section{Multi-Fidelity Gaussian Process}\label{app:mfgp}
In a Gaussian process a function \( \hat{f}(\theta) \) is modeled as:
\[
   \hat{f}(\theta) \sim \mathcal{GP}(\mu(\theta), k(\theta, \theta'))
\]
where \( k(\theta, \theta') \) is the covariance function. Given data \( D_N = \{\Theta, y\} \), the posterior mean and variance at a new point \( \theta_* \) are:
\[
   \mu_* = K(\theta_*, \Theta) K(\Theta, \Theta)^{-1} y
   \]
\[
   \sigma_*^2 = K(\theta_*, \theta_*) - K(\theta_*, \Theta) K(\Theta, \Theta)^{-1} K(\Theta, \theta_*)
   \]

This model represents the joint distribution of multiple fidelities as a multivariate Gaussian process with a specified covariance structure. The covariance matrix in this model includes both correlation terms between fidelities and discrepancy terms within fidelities. Consequently, the HF model \( \hat{f}_H(\theta) \) is expressed in terms of the LF model \( \hat{f}_L(\theta) \) with a discrepancy \( \delta(\theta) \):
   \[
   \hat{f}_H(\theta) = \rho \hat{f}_L(\theta) + \delta(\theta)
   \]
   where \( \rho \) is a scaling factor and \( \delta(x) \) is modeled as a GP. The joint distribution is:
   \[
   \begin{pmatrix}
   \hat{f}_L(x) \\
   \hat{f}_H(x)
   \end{pmatrix} \sim \mathcal{N} \left( 0, \begin{pmatrix}
   K_{LL} & \rho K_{LL} \\
   \rho K_{LL} & \rho^2 K_{LL} + K_{\delta}
   \end{pmatrix} \right)
   \]
For more fidelity levels, the method recursively applies:
   \[
   \hat{f}_{H_i}(x) = \rho_i \hat{f}_{L_i}(x) + \delta_i(x)
   \]

\section{Active Learning Strategy}~\label{subsec:active_learning}
We use integrated variance reduction to quantify and minimize the expected posterior variance over the input space \( \Theta \).  The goal of integrated variance reduction is to minimize the total variance across the design space:
\[
\mathcal{I}(\theta_{\text{new}}) = \int_{\Theta} \sigma^2(\theta \mid \theta_{\text{new}}) \, d\theta
\]
\( \sigma^2(\theta \mid \theta_{\text{new}}) \) is the updated variance at point \( \theta \) after incorporating the information from the new sample \( \theta_{\text{new}} \). The new sampling point \( \theta_{\text{new}} \) is selected by minimizing this integrated variance:
\[
\theta_{\text{new}} = \arg \min_{\theta' \in \Theta} \mathcal{I}(\theta')
\]
In the context of a Gaussian Process model, the integrated variance reduction acquisition function \(\mathcal{I}(\theta)\) simplifies to
\[
\mathcal{I}(\theta)\approx \frac{1}{ \text{N}}\sum_i^{ \text{N}}\frac{k^2(\mathbf{ \theta}_i, \mathbf{ \theta})}{\sigma^2(\mathbf{ \theta})}.
\]
with $\sigma^2(\theta)$ representing the predictive variance at the observed point \(\theta\) and $k$ is the Radial Basis Function (RBF) kernel with $\mathbf{\theta}_i$ sampled randomly. Furthermore, we aim to optimize the acquisition function under parameter constraints—which limit the allowable values of the design parameters—using a gradient descent algorithm to locate the global maximum. These constraints ensure that the parameters remain within feasible ranges, reflecting the practical and structural requirements necessary for maintaining the integrity and functionality of the overall design. The constraints are incorporated into the acquisition function by adding penalty terms that reduce the expected improvement to zero when constraints are violated, discouraging the algorithm from exploring infeasible regions.

\section{Conditional Neural Process}\label{subsec:link_vae}

The core principle of the Conditional Neural Process (CNP) framework is to approximate arbitrary random processes using Gaussian sampling, where the mean $\mu$ and variance $\sigma$ are parameterized by neural networks. In this section, we show that the CNP score $\beta$ can be viewed as an estimate of $t(\boldsymbol{\theta}, \boldsymbol{\phi})$, with the Gaussian distribution representing the posterior of $t(\boldsymbol{\theta}, \boldsymbol{\phi})$. Furthermore, the RED problem can be aligned with the theoretical framework of the Variational Autoencoder (VAE), as detailed in \citet{vae}, with $t(\boldsymbol{\theta},\boldsymbol{\phi})$ interpreted as the latent vector.

We begin by formulating the RED problem within a Bayesian framework: The training data consists of a finite set of \( X_{ki} \) values generated through simulation, and the objective is to construct an encoder \( q \) that approximates the posterior distribution of $t(\boldsymbol{\theta}, \boldsymbol{\phi})$, conditioned on the observed dataset \(\{X_{ki}, \boldsymbol{\phi_{ki}}, \boldsymbol{\theta_{k}}\}\).
\begin{equation}
q(t(\boldsymbol{\theta},\boldsymbol{\phi})|X_{ki},\boldsymbol{\phi_{ki}},\boldsymbol{\theta_k}) \approx p(t(\boldsymbol{\theta},\boldsymbol{\phi})|X_{ki},\boldsymbol{\phi_{ki}},\boldsymbol{\theta_k})
\label{eq:encoder-approx}
\end{equation}
According to Bayes' theorem, the conditioned posterior in Eq.~\ref{eq:encoder-approx} could be calculated with the likelihood of the observed dataset and the prior of $t(\boldsymbol{\theta},\boldsymbol{\phi})$:
\begin{equation}
p(t(\boldsymbol{\theta},\boldsymbol{\phi})|X_{ki},\boldsymbol{\phi_{ki}},\boldsymbol{\theta_k}) \propto \mathcal{L}(X_{ki}|\boldsymbol{\phi_{ki}}, \boldsymbol{\theta_k},t(\boldsymbol{\theta},\boldsymbol{\phi}))p(t(\boldsymbol{\theta},\boldsymbol{\phi}))
\end{equation}
The prior \( p(t(\boldsymbol{\theta}, \boldsymbol{\phi})) \) is conventionally set as a constant. Let $\mathcal{L}_k$ represents the $k$-th simulation $\mathcal{L}_k$, the combined likelihood of the full dataset is therefore:
\begin{equation}
    \mathcal{L}(X_{ki}|\boldsymbol{\phi_{ki}}, \boldsymbol{\theta_k},t(\boldsymbol{\theta},\boldsymbol{\phi})) = \prod_{k=1}^{K} \mathcal{L}_k = \prod_{k=1}^{K} \prod_{i=1}^{N_k}\text{Bernoulli}(x=X_{ki}|p=t(\boldsymbol{\theta_k},\boldsymbol{\phi_{ki}}))
\end{equation}
Note that the ground truth of $t(\boldsymbol{\theta},\boldsymbol{\phi})$ is unknown. In the Bayesian framework, we can only estimate it with a probability density function, which is $q(t(\boldsymbol{\theta},\boldsymbol{\phi}))$. The estimation of the likelihood is therefore:
\begin{equation}
\mathcal{L}(X_{ki}|\boldsymbol{\phi_{ki}}, \boldsymbol{\theta_k},q(t)) =\prod_{k=1}^{M} \prod_{i=1}^{N_k} \int \text{Bernoulli}(x=X_{ki}|p=t(\boldsymbol{\theta_k},\boldsymbol{\phi_{ki}})) q(t(\boldsymbol{\theta},\boldsymbol{\phi})) dt
\label{eq:likelihood}
\end{equation}

It is important to note that the quantity \( t \) here is not a variable but a function, and the integration is performed in the Hilbert space. The problem, therefore, becomes to find the function $q^*(t(\boldsymbol{\theta},\boldsymbol{\phi}))$ so that:
\begin{equation}
    q^* = \arg\min_{q \in f(\boldsymbol{\theta},\boldsymbol{\phi})} \mathcal{L}(X_{ki}|\boldsymbol{\phi_{ki}}, \boldsymbol{\theta_k},q(t))
\end{equation}
Here $f(\boldsymbol{\theta},\boldsymbol{\phi})$ represents the set of arbitrary normalized functions of $(\boldsymbol{\theta},\boldsymbol{\phi})$. While this optimization problem is mathematically solvable, the integration computation in Hilbert space is mathematically non-trivial and computationally expensive, rendering this solution impractical.

Then, here comes the CNP model, which simplifies and tackles this optimization problem in the following steps:

First, approximate $q$ with parameterized Gaussian, which is a natural choice if we regard the task as a statistical parameter estimation for $t(\boldsymbol{\theta},\boldsymbol{\phi})$:
\begin{equation}
    q_{NN}(t(\theta,\phi)) = \mathcal{N}(\mu_{NN}(\boldsymbol{\theta},\boldsymbol{\phi},\boldsymbol{w}), \sigma^2_{NN}(\boldsymbol{\theta},\boldsymbol{\phi},\boldsymbol{w}))|_{X_{ki},\boldsymbol{\phi_{ki}}, \boldsymbol{\theta_k}}
\end{equation}

Then, the neural network is trained to minimize the likelihood described in Eq.~\ref{eq:likelihood}. Therefore, the CNP model actually performs a statistical estimation for $t(\boldsymbol{\theta},\boldsymbol{\phi})$ by approximating the posterior $q$ with Gaussians. 

If we consider $t(\boldsymbol{\theta},\boldsymbol{\phi})$ as the latent vector in the VAE model, the pre-defined Bernoulli process acts as a probabilistic ``decoder'', which generates data given the latent vector. 

\begin{figure}[h]
\begin{center}
    \centering
    \includegraphics[width=\linewidth]{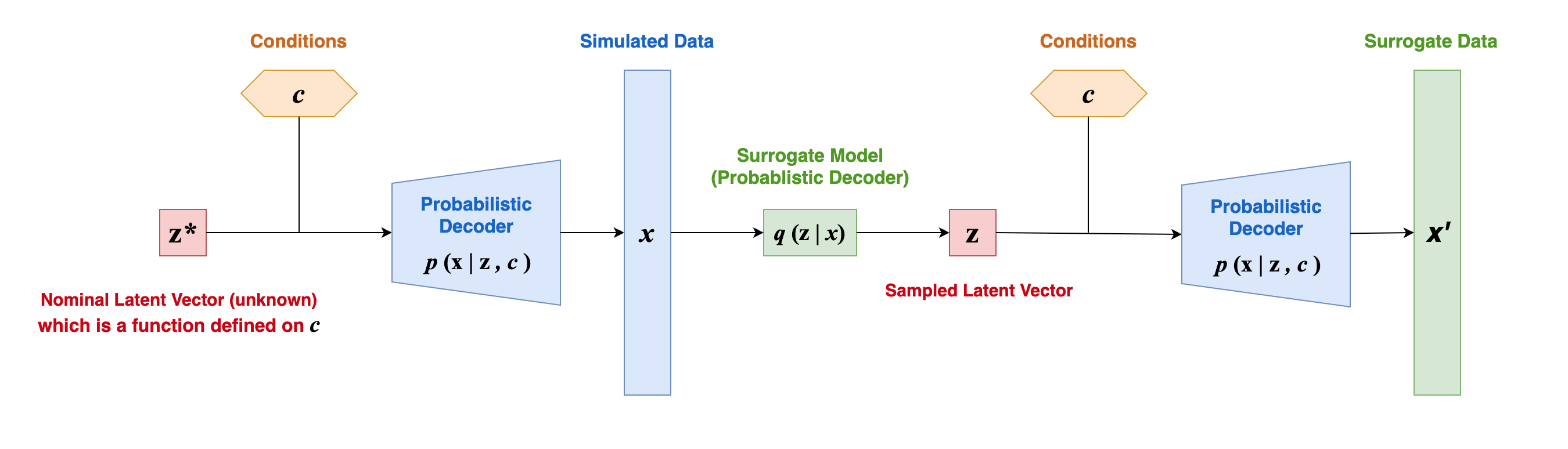}
    \caption{Architecture of Surrogate Model for the RED Problem}
    \label{fig:vae-formalism-diagram}
\end{center}
\end{figure}

With this framework, our task can be generally described as follows: Assume we have a predefined probabilistic decoder, $p(\boldsymbol{x}|\boldsymbol{z},\boldsymbol{c})$, where $\boldsymbol{z}$ represents the latent vector and $\boldsymbol{c}$ is the condition. Additionally, we have a simulation informed by domain knowledge that generates data based on the nominal latent vector $\boldsymbol{z^*}$, though its exact value is unknown. Our objective is to develop a surrogate model that performs statistical estimation, represented as $q(\boldsymbol{z}|\boldsymbol{x})$, to infer the nominal latent vector from the simulated data $\boldsymbol{x}$. The posterior distribution $q(\boldsymbol{z}|\boldsymbol{x})$ serves the same role as the probabilistic encoder in a VAE model. We can then sample the latent vector $\boldsymbol{z}$ from $q(\boldsymbol{z}|\boldsymbol{x})$, which can subsequently be used to generate the "reconstructed" (surrogate) data $\boldsymbol{x'}$. The illustration of this architecture is shown in Figure~\ref{fig:vae-formalism-diagram}.

In this work, as the primary objective is to optimize the design parameters rather than generate additional simulated datasets, we utilize only the averaged CNP score $\beta$ as input to the subsequent MFGP model. Nevertheless, it is important to highlight that the trained CNP model has the potential to generate surrogate data as a fast simulator. This capability can be especially beneficial in other studies where the detailed distribution of the latent vector parameters is of interest, enabling more efficient exploration of the parameter space and supporting applications such as uncertainty quantification and model validation.

For the RED problem, specifically, \( t(\boldsymbol{\theta}, \boldsymbol{\phi}) \) corresponds to the latent vector \( \boldsymbol{z} \), but with infinite dimensions, as it represents a continuous function over \( (\boldsymbol{\theta}, \boldsymbol{\phi}) \). $(\boldsymbol{\theta},\boldsymbol{\phi})$ acts as the condition $\boldsymbol{c}$, and the fusion of latent vector and condition is performed by plugging $\boldsymbol{\theta}, \boldsymbol{\phi}$ into the function $t$ to get $t(\boldsymbol{\phi},\boldsymbol{\theta})$

It is worth noting that the latent vector, denoted by $\boldsymbol{z}$, in our model, is not a conventional finite-dimensional vector but rather a vector situated in a Hilbert space. Dimensionality reduction can be achieved through quantization, which involves discretizing the space by creating a grid over $(\boldsymbol{\theta}, \boldsymbol{\phi})$ and computing values only at the selected grid points. An alternative approach is to project the latent vector onto a set of basis functions and impose a cutoff. For instance, with Fourier decomposition, the vector's coefficients can be retained up to a certain frequency limit, effectively serving as a high-frequency cutoff. In this work, we adopt the CNP model to represent it with a \textit{fixed} dimensionality.

\section{Comparative Analysis of Multi-fidelity Approaches}\label{subsec:mf_comparsion}

\begin{figure}[hbt!]
    \centering
    \includegraphics[width=0.98\linewidth]{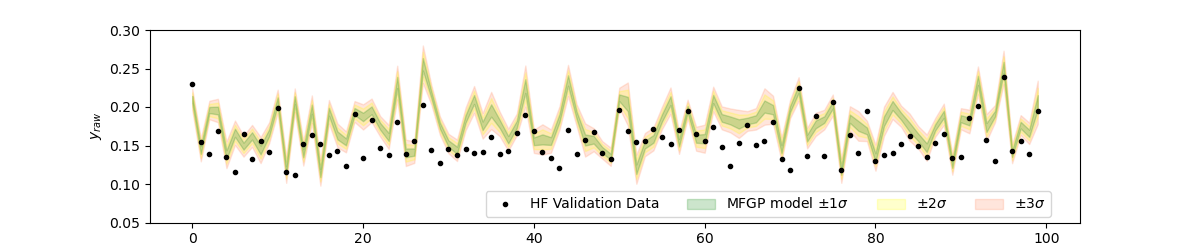}
    \caption{Validation of the simplified MFGP model using only LF \(y_{Raw}\) and HF \(y_{Raw}\) data, excluding the CNP. The MFGP model fails to adequately describe \(y_{Raw}\), as demonstrated by its overly narrow prediction bands and poor alignment with the HF validation data. Despite the narrow uncertainty bands, the predictions lack physical relevance, highlighting the model’s inability to capture the complexity of the design space without the CNP}
    \label{fig:mf-validation}
\end{figure}

Through the development and application of RESuM, we demonstrated the significant benefits of employing a multi-fidelity model enhanced by the CNP to solve RED problems. RESuM integrates three different fidelity levels, two of which are generated using the CNP. When compared to a simpler multi-fidelity Gaussian Process (MFGP) model that only utilized the LF \(y_{Raw}\) and HF \(y_{Raw}\) of the exact same LF and HF data, excluding the CNP outputs \(y_{CNP}\), the contrast in outcomes was striking. The model without the CNP outputs produced predictions that lacked physical relevance. The simplified model not only failed to capture the complex dependencies between design parameters, but its predictions were also physically inconsistent. As shown in Figure~\ref{fig:mf-validation}, the prediction bands for the simplified MFGP model are excessively narrow and fail to capture the actual variability of the HF validation data. The model's inability to describe \(y_{Raw}\) is evident, as it does not adequately reflect the complex interactions within the design space. The model's 1 \(\sigma\), 2 \(\sigma\), and 3 \(\sigma\) confidence intervals are unrealistically tight, offering little insight into the true uncertainties of the system. With that, the coverage at 1 \(\sigma\), 2 \(\sigma\) and 3 \(\sigma\) for the simplified model was only 12\%, 24\% and 47\%, a notably poor result compared to the RESuM model, which achieved a 1 \(\sigma\), 2 \(\sigma\) and 3 \(\sigma\) coverage of 69\%, 95\% and 100\% with much more realistic uncertainty predictions (compare Figure~\ref{fig:validation}). Similarly, attempts to use Gaussian regression with the 100 HF validation samples alone proved insufficient to model the full 5-dimensional design parameter space. The Gaussian regression model struggled to generalize across all design dimensions, particularly failing to produce meaningful predictions when all parameters were included. 

These findings underscore the complexity of the design space and the limitations of traditional Gaussian regression or simpler multi-fidelity approaches with limited HF data. In contrast, the CNP-enabled RESuM model effectively reduced statistical variance and provided meaningful physical insights by capturing complex relationships between design parameters \(\theta\) and event-specific parameters \(\phi\). Specifically, the CNP outputs \(y_{CNP}\) smoothed the predictive landscape, allowing the multi-fidelity Gaussian Process to interpolate and extrapolate across the design space with much lower uncertainty. The simplified MFGP model and Gaussian regression both failed to capture consistent dependencies between the design parameters and the neutron background reduction, resulting in physically implausible predictions. The inclusion of the CNP was therefore critical in modeling the full 5-dimensional design parameter space, enabling RESuM to achieve a substantial reduction in computational resources while still providing accurate and robust predictions for optimizing the neutron moderator design in the LEGEND experiment.

These observations reinforce the conclusion that the CNP framework is essential for generating meaningful, accurate predictions in the RESuM model. By reducing uncertainty and improving physical relevance, the CNP-based approach significantly enhances the model's utility for optimizing complex design problems. The ultimate reduction in computational resources, combined with more precise predictions, underlines the effectiveness of this multi-fidelity approach.

\end{document}